\documentclass[aps,prb,twocolumn,groupedaddress,floatfix,showpacs]{revtex4} 
\usepackage{graphics}
\usepackage{subfigure}
\usepackage{epsfig}
\usepackage{dcolumn}
\usepackage{bm}
\usepackage{overpic} 

\begin{document}

\title{Composition and structure of Pd nanoclusters in SiO$_x$ thin film}

\author{Annett Th\o gersen}
\author{Jeyanthinath Mayandi}
\author{Lasse Vines}

\affiliation{Centre for Materials Science and
  Nanotechnology, University of Oslo, P.O.Box 1126 Blindern, N-0318 Oslo, Norway}

\author{Martin F. Sunding}
\author{Arne Olsen}
\affiliation{Department of Physics, University of Oslo, P.O.Box 1048 Blindern, N-0316 Oslo, Norway}

\author{Spyros Diplas}
\affiliation{SINTEF Materials and Chemistry, P.B 124 Blindern, N-0314 Oslo, Norway and Centre for Material Science and Nanotechnology, University of Oslo.}

\author{Masanori Mitome}
\author{Yoshio Bando}
\affiliation{National Institute of Material Science, Namiki 1-1, Tsukuba, Ibaraki, 305-0044 Japan}

\date{\today}

\begin{abstract}
The nucleation, distribution, composition and structure of Pd nanocrystals in SiO$_2$ multilayers containing Ge, Si, and Pd are studied using High Resolution Transmission Electron Microscopy (HRTEM) and X-ray Photoelectron Spectroscopy (XPS), before and after heat treatment. The Pd nanocrystals in the as deposited sample seem to be capped by a layer of PdO$_x$. A 1-2 eV shift in binding energy was found for the Pd-3d XPS peak, due to initial state Pd to O charge transfer in this layer. The heat treatment results in a decomposition of PdO and Pd into pure Pd nanocrystals and SiO$_2$. 
\end{abstract}

\maketitle

\section{Introduction}

\noindent Material systems containing silicon and germanium nanocrystals have attracted much attention due to their optical and electronic properties \cite{agan:intro, salh:intro}, as well as their potential applications in photo detectors \cite{agan:1}, light emitters \cite{agan:2}, single electron transistors \cite{agan:3} and non-volatile memories \cite{kanoun:intro}. Nanoclusters embedded in a SiO$_2$ matrix is an attractive option towards nanocluster based device development \cite{yang:intro}. The most important factors influencing the optical properties of the SiO$_2$-nanocluster devices are size, spatial distribution, atomic and electronic structure as well as the surface properties of the nanoclusters. 

Pd is an especially interesting material when used for catalytic converters in auto-mobile technology for the elimination of NO$_X$ (Nitrogen Oxides) in the exhaust gases of gasoline engines \cite{Pd:7}. Considerable research has also been conducted in the use of Pd catalysts for the combustion of methane. Particle morphology and oxidation state can play an important role in defining the active sites on Pd catalysts. Pd is also found in other applications such as Granular Metal (GM) films, cermets or nano-cermets, where metal particles on MgO cubes \cite{Pd:sio2}. Transition metal particles have interesting properties due to quantum size effect, owing to the dramatic reduction of the number of free electrons \cite{Miyake:pd}. The nanoparticle  matrix may formed advanced material system with new electronic, magnetic, optic, and thermal properties \cite{Miyake:pd}.

In this work, samples containing both Pd nanocrystals and Ge clusters embedded in SiO$_2$ layers supersaturated with Si, were studied in detail, before and after heat treatment. The formation, composition, distribution, and the atomic and electronic structure of Ge and Pd nanoclusters were studied by High Resolution Transmission Electron Microscopy (HRTEM) and X-ray Photoelectron Spectroscopy (XPS).

\section{Experimental}

\noindent The samples were produced by growing a $\sim$3 nm layer of SiO$_2$ on a p-type Si substrate by Rapid Thermal Oxidation (RTO) at 1000$^\circ$C for 6 sec. Prior to growing the RTO layer the wafers were cleaned using a standard RCA procedure (Radio Corporation of America, industry standard for removing contaminants from wafers) followed by immersion in a 10 \% HF solution to remove the native oxide. Then a $\sim$10 nm layer of silicon rich oxide (46 at. \%) was sputtered from a SiO$_2$:Si composite target onto the RTO-SiO$_2$ and heat treated in a N$_2$ atmosphere at 1000$^\circ$C for 30 min, as described in our previous article \cite{Annett:senere, Annett:1}. A $\sim$20nm SiO$_2$ layer containing 0.5 at. \% Ge and 0.5 at. \% Pd, was then sputtered. This sample is reffered to as the as deposited sample (sample \textsf{ASD}) The heat treated sample (sample \textsf{HT}) was then annealed again at 900$^\circ$C in a N$_2$ atmosphere for 30 minutes to nucleate Ge and Pd nanocrystals. Cross-sectional TEM samples were prepared by ion-milling using a Gatan precision ion polishing system with a 5 kV gun voltage.

The nanocrystals were studied with HRTEM using a 300 keV JEOL 3100FEF TEM with an Omega imaging filter. Additional HRTEM images were acquired using a 200 keV JEOL 2010F TEM. XPS was performed in a KRATOS AXIS ULTRA$^{DLD}$ using monochromatic Al K$\alpha$ radiation ($h\nu=1486.6$ eV) on plane-view samples using 0$^\circ$ angle of emission (vertical) and charge compensation with ion energy electrons from a flood gun. The X-ray source was operated at 10 mA and 15 kV. The inelastic mean free path of the Ge$_{3d}$ electrons in SiO$_2$ is $\sim$3.9 nm \cite{xray:ref, nist:1}. Emission normal to the surface results in a photoelectron escape depth of 11.8 nm \cite{nist:1}, which allowed us to study the silicon nanoclusters located 11.8 nm below the surface of the oxide. Accordingly the inelastic mean free path of the Pd-3d electrons in SiO$_2$ is 3.38 nm, which allows us to study Pd nanocrystals within an analysis depth of 10.1 nm. The spectra were peak fitted using the CasaXPS program \cite{casa:xps} after subtraction of a Shirley type background. The spectra were calibrated by adjusting peak positions of the O-1s and Si-2p signals from SiO$_2$ at 533 eV and 103.6 eV respectively, and the Si-2p from the Si substrate at 99.5 eV\cite{xray:ref}. Composition depth profile was performed with Ar ion etching, on a 3x3 mm spot area, at 2 kV, 100 $\mu$A current and a cycle time of 20 s.

\section{Results and Discussion}

\begin{figure}
  \begin{center}
    \includegraphics[width=0.4\textwidth]{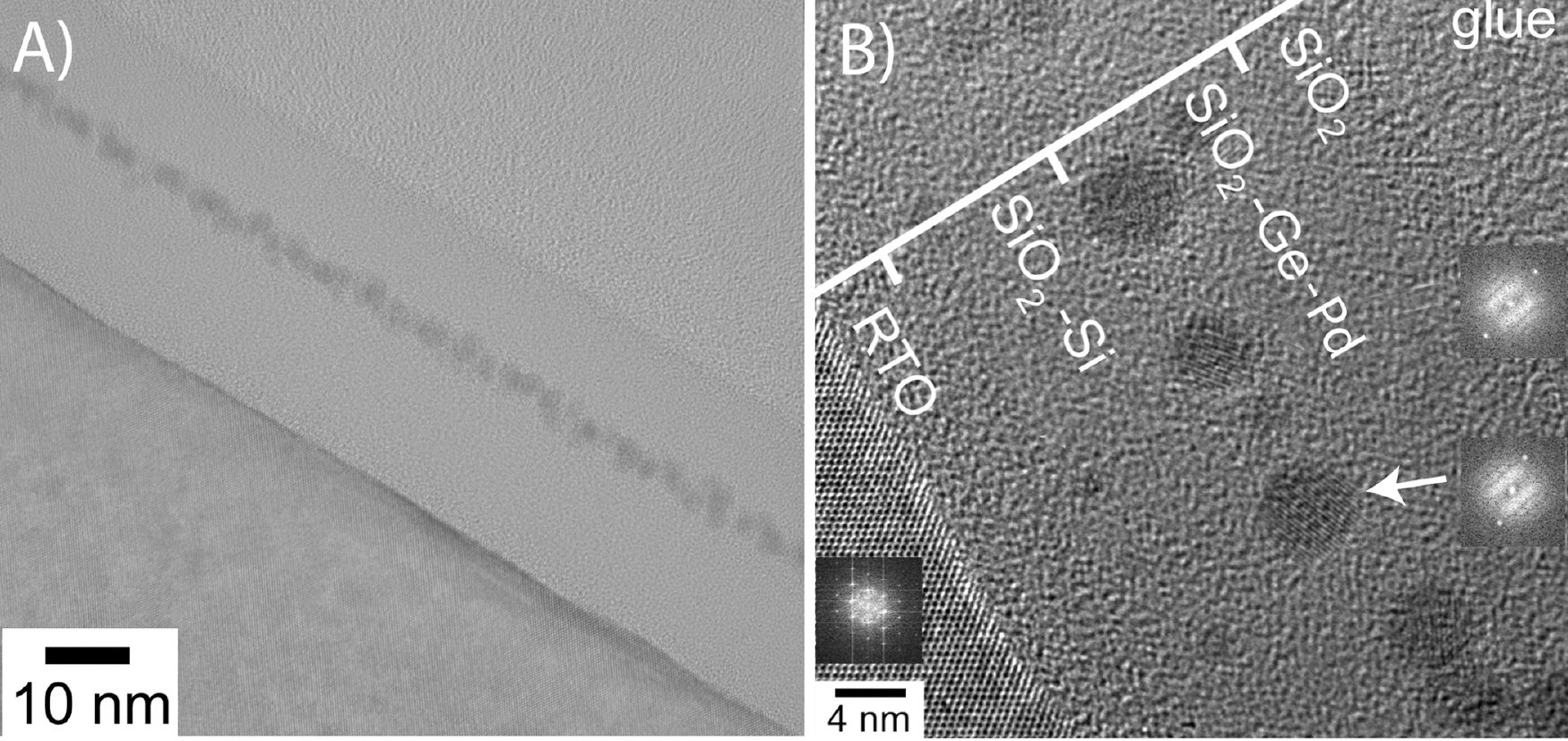}
   \caption{HRTEM images of A) sample \textsf{ASD} and B) \textsf{HT}. The arrows show both crystalline (in the SiO$_2$-Ge layer) and amorphous (in the SiO$_2$-Si-layer) nanoclusters of Pd. Fast Fourier Transform patterns of the nanocrystals are presented as insets and shows that the clusters are crystalline.}
    \label{figure:1}
  \end{center}
\end{figure}

Nanoclusters of different sizes in various layers were observed in both the as deposited (\textsf{ASD}) and heat treated sample (\textsf{HT}), see Figure \ref{figure:1}. Dark Ge and Pd nanoclusters are visible in the SiO$_2$-Ge-Pd layer, 10 nm from the outermost surface (SiO$_2$/glue interface) and 13 nm from the Si substrate. Lattice fringes can be seen in the largest nanoclusters, indicating that many of them are crystalline. The average nanocluster size for sample \textsf{HT} and \textsf{ASD} is 4 nm and 2.5 nm, respectively . Smaller (1-2 nm) clusters are visible in the SiO$_2$-Si layer 6 nm from the Si substrate in sample \textsf{HT}. No lattice fringes were observed in these nanoclusters, indicating an amorphous structure. Since Si has almost no contrast when embedded in SiO$_2$, these nanoclusters contain either Pd or Ge.

\begin{figure*}
  \begin{center}
    \includegraphics[width=1.0\textwidth]{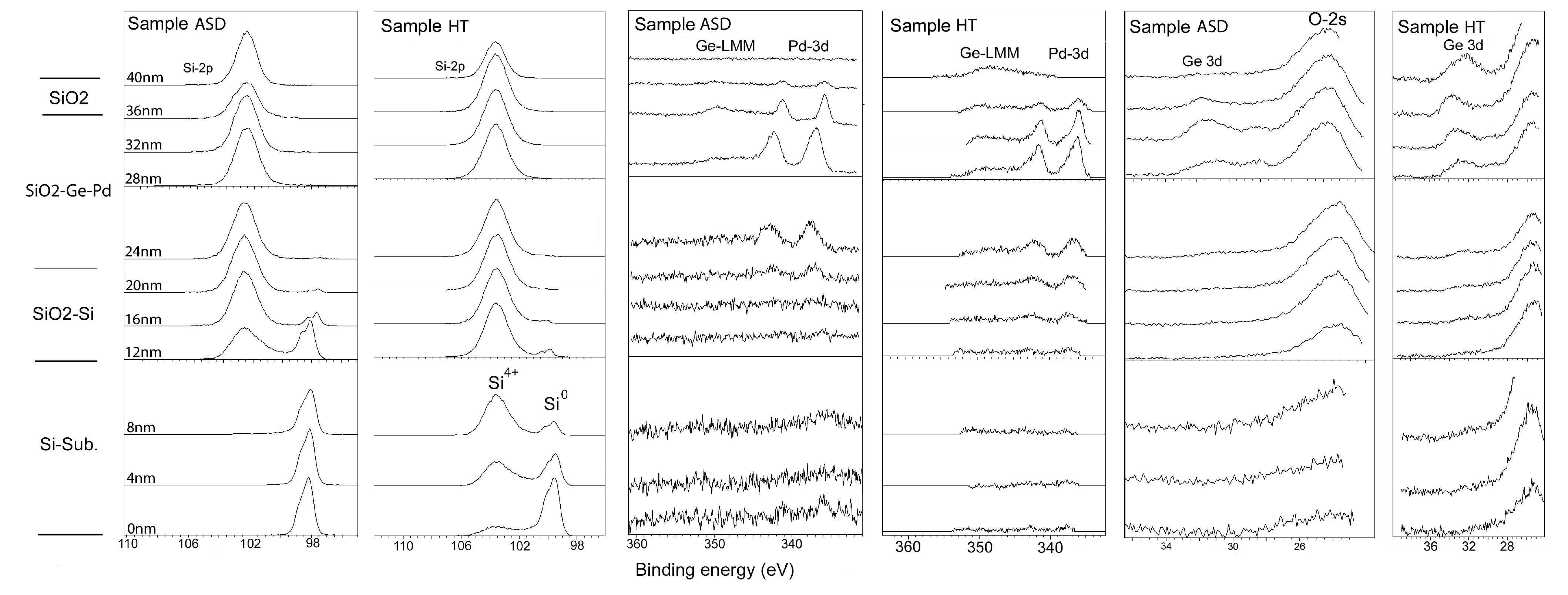}
   \caption{XPS depth profiles of the Si$_{2p}$-, Ge$_{3d}$-, and Pd$_{3d}$-peak of sample \textsf{HT} and \textsf{ASD}.}
    \label{figure:2}
  \end{center}
\end{figure*}

Figure \ref{figure:2} shows the high resolution Si$_{2p}$, Ge$_{3d}$, and Pd$_{3d}$ XPS spectra acquired during depth profiling with Ar$^+$ sputtering of samples \textsf{HT} and \textsf{ASD}. The XPS spectra of sample \textsf{HT} show that Ge is distributed wider in the SiO$_2$-Ge-Pd layer, compared to sample \textsf{ASD}, where Ge is located in a narrow band. In addition, Ge surface segregation seems to occur upon annealing in accordance with previous work by Agan et al. \cite{agan:intro} and Marstein et al. /cite{marstein:1}. Sample \textsf{ASD} looks very similar to sample \textsf{HT}, apart from the location of the Si and Ge (see Figure \ref{figure:2}). The Pd distribution is almost the same in both samples. As seen in Figure \ref{figure:1}, small nanoclusters were observed in the SiO$_2$-Si layer of samples \textsf{HT}. The Ge-3d and Pd-3d spectra of sample \textsf{HT} show very little or no increase in the Ge and Pd concentration in the SiO$_2$-Si layer. This indicates that the nanoclusters shown in Figure \ref{figure:1} that are present in the SiO$_2$-Si layer, have diffused to this area during electron beam exposure. There is a visible shift in the position of the Pd peaks with the depth in sample \textsf{ASD}. This shift, not seen in sample \textsf{HT}, will be discussed in Section \ref{sec:core-shift}. Figure \ref{figure:3} shows the elemental distribution with depth for sample \textsf{ASD} and \textsf{HT}. The quantification is based upon peak fitting, being shown in more  in the following sections.

The results of the as deposited samples are more complicated to interpret. We therefore present first the less complicated results of the heat treated samples, followed by the as deposited ones.

\section{Results and Discussion: Heat treated samples}
\label{heat}

The composition of the heat treated samples measured with XPS, EELS, EFTEM and EDS is presented in Section \ref{sec:comheat}, while the atomic structure found by TEM is presented in Section \ref{atomicstructure}. 

\subsection{Nanocluster composition}
\label{sec:comheat}

\begin{figure}
  \begin{center}
    \includegraphics[width=0.5\textwidth]{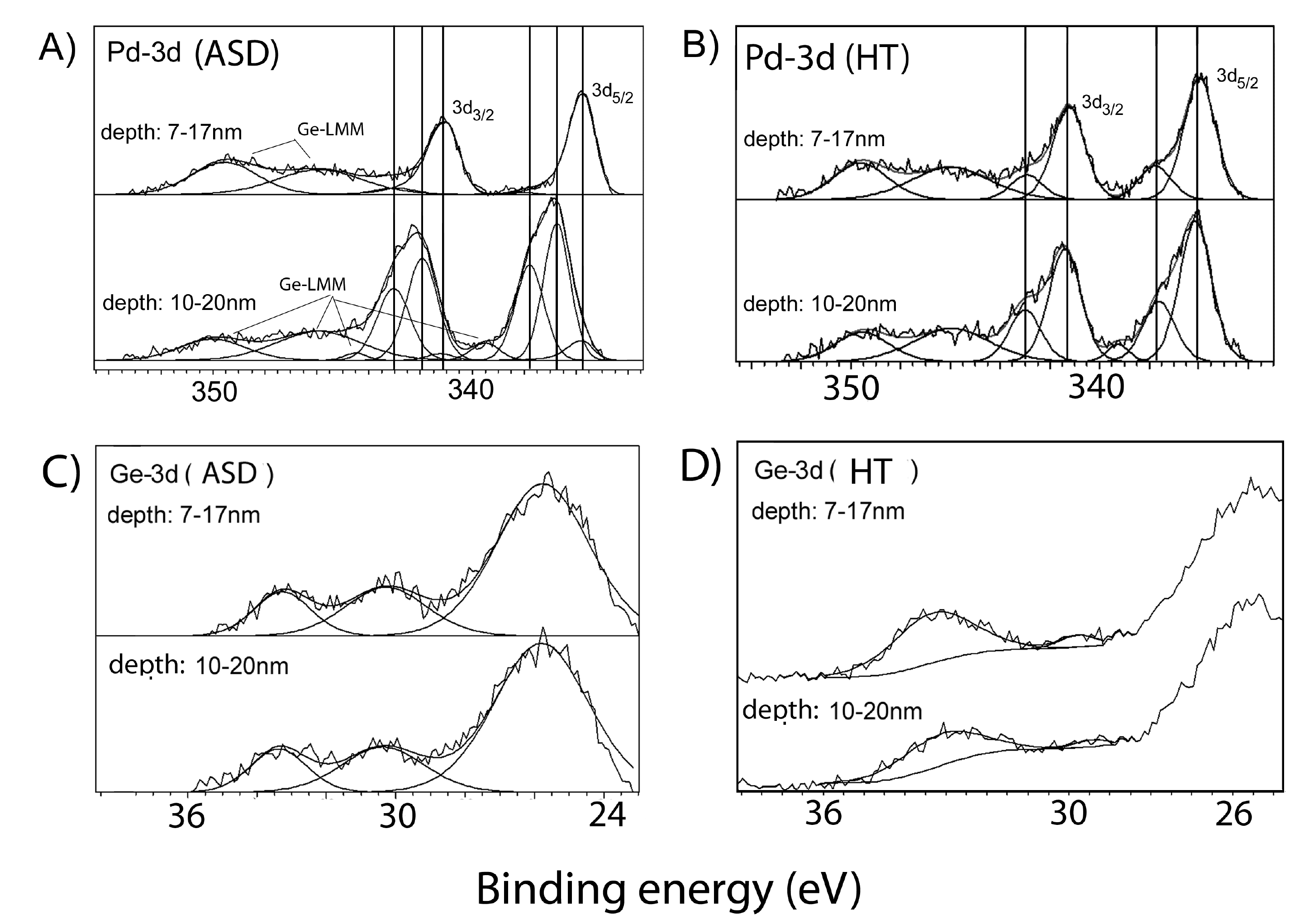}
   \caption{XPS spectra of A) the Pd-3d peak of sample \textsf{ASD}, B) the Pd-3d peak of sample \textsf{HT}, C) the Ge-3d peak of sample \textsf{ASD}, and D) the Ge-3d peak of sample \textsf{HT}. The detection depth is with respect to the sample surface.}
    \label{figure:6}
  \end{center}
\end{figure}

XPS was used to identify the chemical state of different elements present in sample \textsf{HT}. Figure \ref{figure:6}B and D shows the fitted XPS spectrum of the Pd-3d and Ge-3d peak. The measured binding energies of the two Ge peaks were 29.8 eV and $33.1 \pm 0.2$ eV, with an energy separation $\Delta E=3.3$ eV. The literature binding energy value for pure Ge is 29.4 eV \cite{xray:ref} and 32.5 eV for GeO$_2$ \cite{wu:ge}. The Ge-3d energy separation/chemical shift between Ge$^0$ and GeO$_2$ is reported to be 3.3 eV \cite{nist:1}. The chemical shift is less susceptible to energy referencing when the same reference is used and both peaks defining the chemical shift exist in the same spectrum. Therefore the above data suggests that the peak at 29.8 corresponds to pure Ge, while the peak at 33.1 $\pm 0.2$eV and $33.3 \pm 0.2$ eV is from GeO$_2$.

The XPS spectrum in Figure \ref{figure:6}B shows the Pd-3d peak. The spectrum is fitted with six components. The two peaks located at binding energies of 348.9 eV and 345.9 eV belong to Ge-LMM. The remaining four peaks may be contributions from Pd$^0_{3d}$ and Pd$^{2+}_{3d}$. The measured binding energies are presented in Table \ref{table:Pd}, and the chemical shifts in Table \ref{table:PdPd}. Pure Pd has a binding energy of 340.4 eV (3d$_{3/2}$) and 335.1 eV (3d$_{5/2}$) \cite{xray:ref}, PdO of 341.6 eV (3d$_{3/2}$) and 336.3 eV (3d$_{5/2}$), and PdO$_2$ of 343 eV (3d$_{3/2}$) and 337.9 eV (3d$_{5/2}$) \cite{kim:pd}. The reference values show chemical shifts for the Pd$^{2+}$-Pd$^0$=1.2 eV, and for Pd$^{4+}$-Pd$^0$=2.6 eV.  

The two smallest peaks in Figure \ref{figure:6}B at 341.6 eV and 336.3 eV fit well with PdO, while the larger peaks at 340.1 eV and 334.8 eV correspond to pure Pd. This is in agreement with the chemical shift of 1.5 between the Pd$^{2+}$-Pd$^0$ of both the 3d$_{3/2}$ and 3d$_{5/2}$ peaks. The Pd$^0$ peaks were fitted using slightly asymmetric components, with a FWHM of about 1 eV. These XPS results are shown and discussed in Section \ref{sec:core-shift} and \ref{sec:shiftasd}.

\begin{figure}
  \begin{center}
    \includegraphics[width=0.5\textwidth]{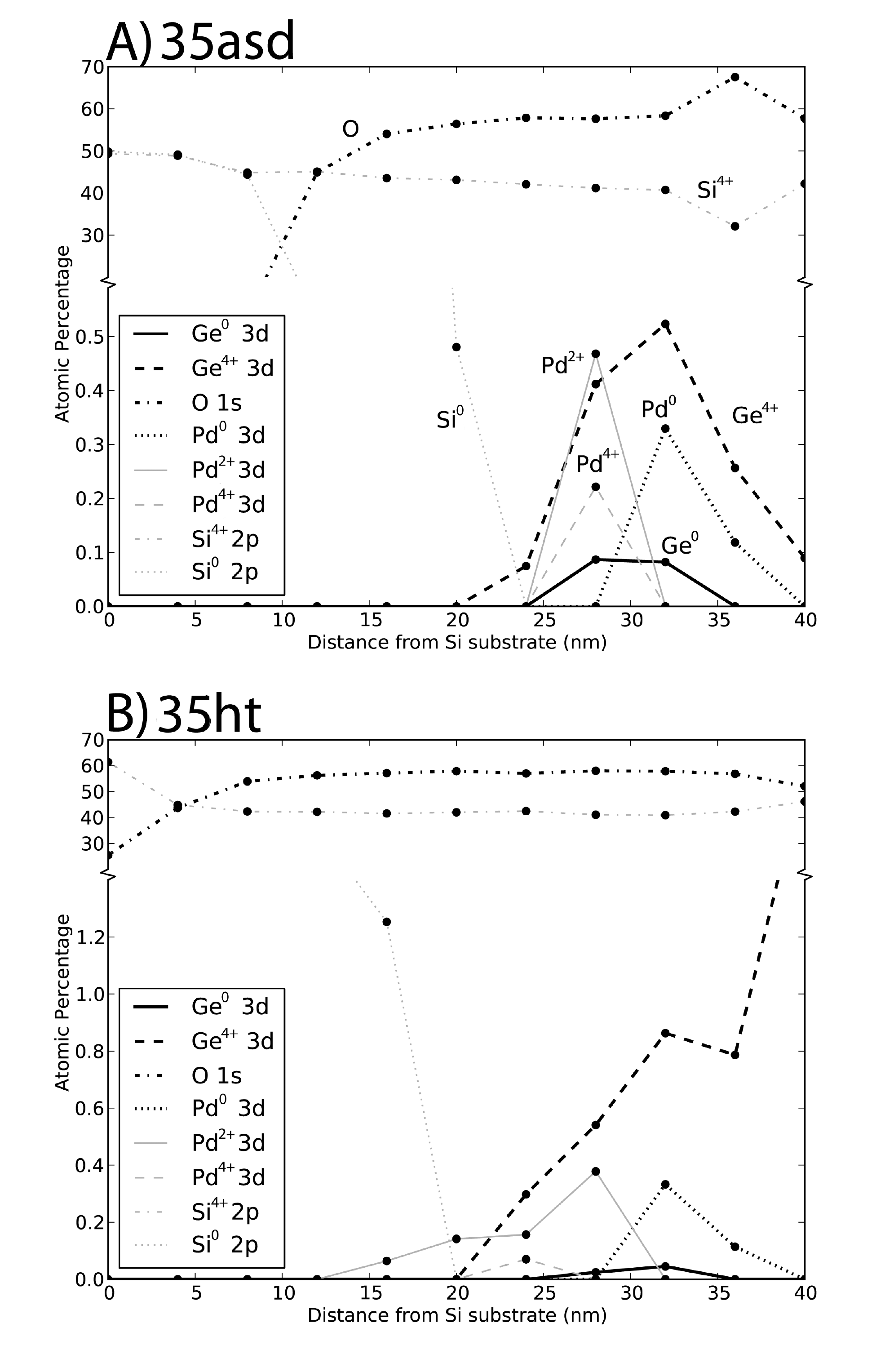}
   \caption{Elemental profile obtained from the XPS depth profile spectra of A) sample \textsf{ASD}, and B) sample \textsf{HT}.}
    \label{figure:3}
  \end{center}
\end{figure}

Figure \ref{figure:2} shows significant differences in the Si$^{4+}$/Si$^0$ intensity ratio in the SiO$_2$-Si layer between the two samples. The increased Si$^{4+}$/Si$^0$ intensity ratio in the SiO$_2$-Si zone as well as in the Si substrate of the annealed sample is attributed to oxidation upon annealing. This oxidation may also be responsible for the small decrease in pure Ge concentration, as seen in Figure \ref{figure:3}.

\subsection{Atomic structure of the nanocrystals}
\label{atomicstructure}

\begin{table}[h!]
\caption{The observed d-values found from the FFT patterns compared to reference d values.}
\begin{tabular}{llllll}
\hline
\hline
Sample & Observed  & Pd \cite{pd:1} & Pd$_3$O$_4$ \cite{Pdref:1} & PdO \cite{Pdref:2} & Pd$_2$Si \cite{ps2si:ref}  \\
& d-value & (nm) & (nm) & (nm) & (nm) \\
& (nm) &  & &  &  \\
\hline
\textsf{HT} & 0.198 & 0.195  & & &  \\
\textsf{ASD} & 0.226 & 0.224 & & & \\
\textsf{ASD} & 0.210 & & 0.204 & 0.211 & 0.211 \\
\hline
\hline
\end{tabular}
\label{table:dvalues}
\end{table}

The atomic structure of the nanocrystals in sample \textsf{HT} were studied by Fast Fourier Transform (FFT) patterns. FFT patterns of the Si substrate and two nanocrystals are presented as insets in Figure \ref{figure:1}. The FFT of the Si substrate is used as a reference, since the Si unit cell dimensions are known ($d=0.543$ nm). The FFT of the two nanocrystals yielded a d-value of $0.198 \pm 0.001$ nm. The HRTEM image shown in Figure \ref{figure:1} shows a squared pattern with the same lattice plane spacing of $d=0.198$ nm. The measured d-values as well as references are presented in Table \ref{table:dvalues}. Pure Pd has a Fm-3m space group, with lattice parameter 0.389 nm. The largest d-value is (200): 0.195nm \cite{pd:1}. This fits well with the measured d-values.

\begin{figure}
  \begin{center}
    \includegraphics[width=0.4\textwidth]{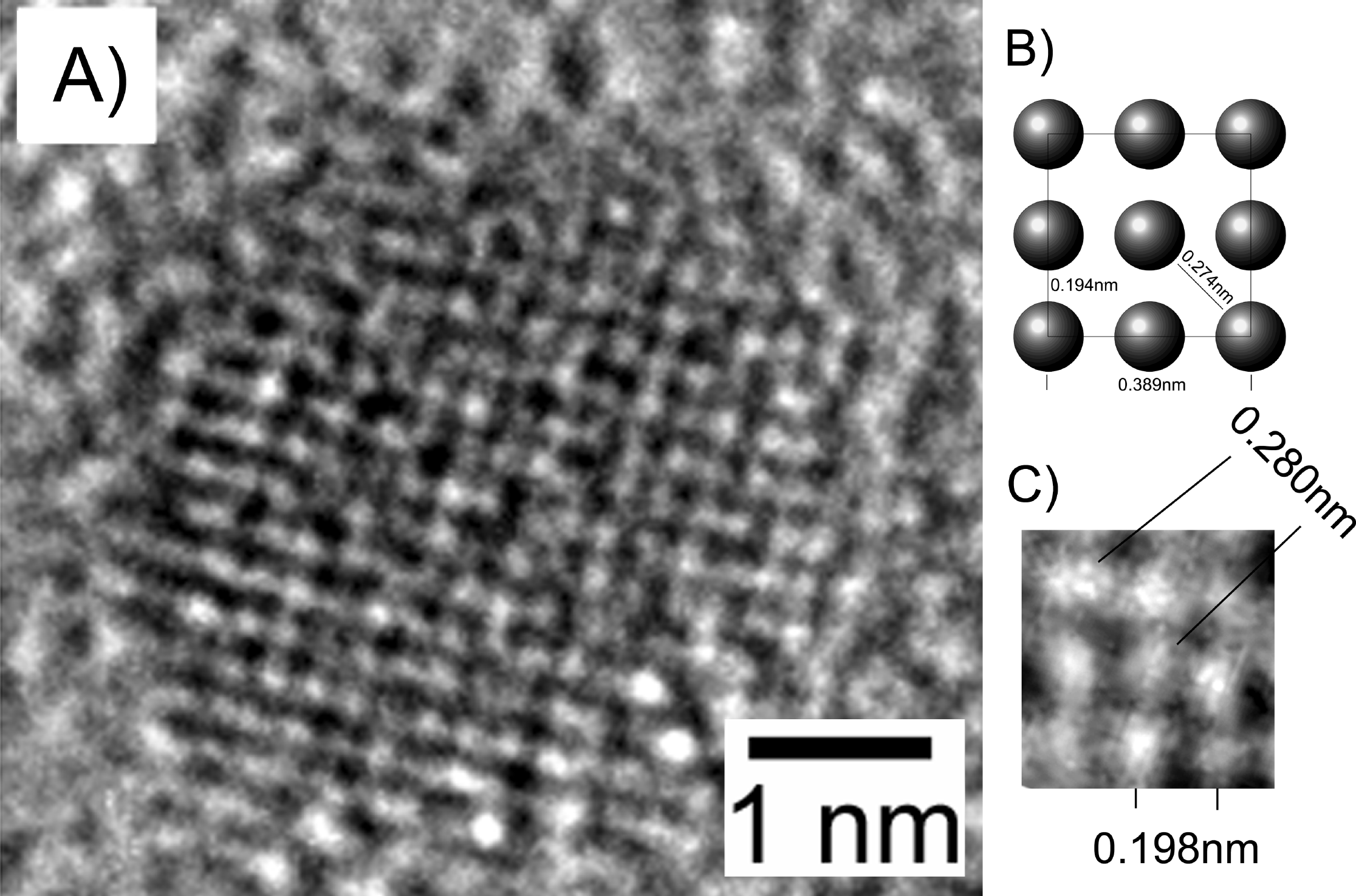}
   \caption{A) HRTEM image of a Pd nanocrystal in sample \textsf{ASD}, B) a sketch of the (100) zone axis of a Pd nanocrystal, and C) a higher magnified image of the atomic structure of the nanocrystal.}
    \label{figure:4}
  \end{center}
\end{figure}

\section{Results and Discussion: As deposited samples}
\label{asdeposited}

The composition and atomic structure of sample \textsf{ASD} is presented in the following sections. XPS measurements of sample \textsf{ASD} is presented in Section \ref{sec:core-shift}, and the discussion of the results in in Section \ref{sec:shiftasd}.

\begin{figure}
  \begin{center}
    \includegraphics[width=0.4\textwidth]{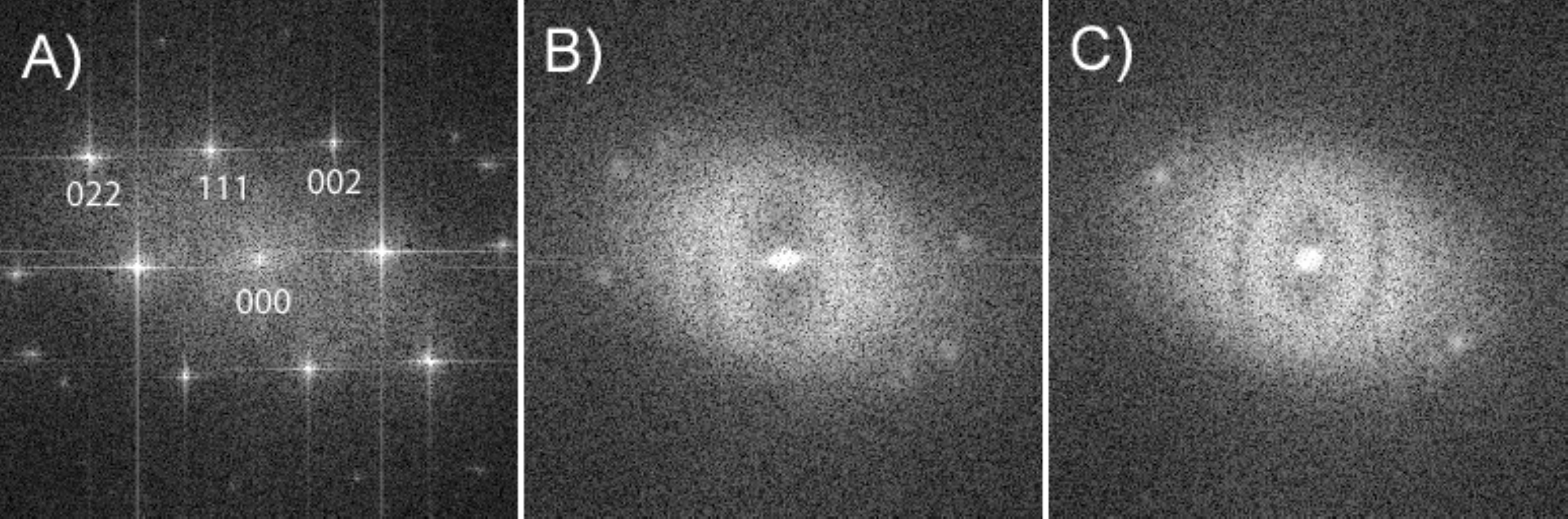}
   \caption{FFT patterns from A) the Si substrate, and B) and C) are from two Pd nanocrystals in sample \textsf{ASD}.}
    \label{figure:5}
  \end{center}
\end{figure}

\subsection{Nanocluster composition and atomic structure}
\label{nanoclustercomp}

A HRTEM image of a Pd nanocrystal in sample \textsf{ASD} is presented in Figure \ref{figure:4}. FFT patterns from the Si substrate and from two nanocrystals are presented in Figure \ref{figure:5}. FFT patterns of three different nanocrystal areas in sample \textsf{ASD} yielded the measured d-values of $0.221 \pm 0.002$ nm, $0.226 \pm 0.002$ nm, and $0.210 \pm 0.002$ nm. The experimentally observed d values are then compared to reference values (presented in Table \ref{table:dvalues}). The d-values measured on the three nanocrystal areas in sample \textsf{ASD} fit with Pd$_3$O$_4$, PdSi, Pd$_2$Si, PdO, and pure Pd.

Voogt et al. \cite{voogt:pd} studied PdO particles with a metallic Pd core. The surface tension of PdO is lower than the surface tension of pure metallic Pd, and upon annealing Pd will not dissolve in PdO \cite{voogt:pd}. It is therefore reasonable to find particles containing, a metallic core and an oxide skin when heat treated at higher temperatures, because the migration of atoms is relatively easy. During heat treatment this core will grow linearly proportional to the surface area, transforming to Pd and SiO$_x$ towards reaching thermodynamic equilibrium. This may explain why we see more Pd oxide in the as deposited samples. 

It is therefore reasonable for the Pd nanocrystals in the as deposited samples to have a few atomic layers of PdO and/or Pd$_2$Si around them. During heat treatment, these compounds will react to form pure Pd and SiO$_2$. XPS results of these samples is shown and discussed in the next section.

The HRTEM and XPS of the as deposited and heat treated sample shows that Ge and Pd behave in an opposite manner. After annealing, Pd decomposes into pure Pd and SiO$_2$, whereas Ge oxidises. This may be influenced by residual oxygen in the annealing ambient. The differences in oxidation may also be due to oxygen transfer from Pd to Ge (and Si), as a result of differences in the enthalpy.

\subsection{Pd-3d binding energy shifts}
\label{sec:core-shift}

\begin{table}[h!]
\caption{The binding energy peak positions in sample \textsf{ASD} and \textsf{HT}}
\begin{tabular}{llll}
\hline
\hline
Peak & Sample \textsf{ASD} & Sample \textsf{HT} & References \footnotemark[1]\footnotemark[2]\\
 & E$_B$ (eV) & E$_B$ (eV) & E$_B$ (eV)\\
\hline
Pd$^0$-3d$_{3/2}$ & 341.0 & 340.1 & 340.4\\
Pd$^0$-3d$_{5/2}$ & 335.7 & 334.8 & 335.1\\
\hline
Pd$^{2+}$-3d$_{3/2}$ & 341.9 & 341.6 & 341.6\\
Pd$^{2+}$-3d$_{5/2}$ & 336.8 & 336.3 & 336.3\\
\hline
Pd$^{4+}$-3d$_{3/2}$ & 343.0 & & 343.0\\
Pd$^{4+}$-3d$_{5/2}$ & 337.9 & & 337.9\\
\hline
\hline
\end{tabular}
\footnotetext[1]{Reference \cite{xray:ref}}
\footnotetext[2]{Reference \cite{kim:pd}}
\label{table:Pd}
\end{table}

The Pd-3d and Ge-3d spectra from sample \textsf{ASD} are shown in Figure \ref{figure:6}A and C, while the Pd-3d spectra from sample \textsf{HT} are shown in Figure \ref{figure:6}B and D. The experimental binding energies, chemical shifts and reference values are shown in Table \ref{table:Pd} and Table \ref{table:PdPd}. We observe a shift in the Pd-3d binding energy with depth in the as deposited sample compared to the heat treated one (Figure \ref{figure:6}). The chemical shift is defined as the binding energy difference 3d$^{2+/4+}_{3/2}$- 3d$^0_{3/2}$. HRTEM image of samples \textsf{ASD} shown in the previous section gave d-values which match well with Pd$_3$O$_4$ (Pd$^{+4}$ and Pd$^{+2}$), PdSi (Pd$^{+4}$), Pd$_2$Si (Pd$^{+2}$), PdO (Pd$^{+2}$), and pure Pd (Pd$^0$).

The Pd-3d peak for sample \textsf{ASD} is visible at sample depth range of 7-17 nm and 10-20 nm, due to a 10 nm photoelectron escape depth for Pd in SiO$_2$. The depth was determined using the Si-2p peak from the Si substrate and the photoelectron escape depth. The spectrum at 7-17 nm is from the top of the band of Pd nanocrystals, and the spectrum at 10-20 nm is from the bottom.

The Pd-3d spectra of both samples at a depth of 7-17 nm show four peaks in addition to the peaks belonging to Ge-LMM. Figure \ref{figure:2} shows the Si-2p peaks of sample \textsf{ASD}. Only Si from SiO$_2$ (Si$^{4+}$) was detected at either depths. This in combination with the Pd peaks means that Pd$_2$Si and PdSi are not present in the samples. The smaller peaks in Figure \ref{figure:6} A must therefore result from PdO and/or Pd$_3$O$_4$.

The chemical shift between the two 3d$_{3/2}$ and 3d$_{5/2}$ peak maxima in Figure \ref{figure:6} A at a depth of 7-17 nm are 2 eV and 2.2 eV respectively for sample \textsf{ASD}. The chemical shift values are higher than the expected one for Pd$^{2+}$ (1.2 eV) and lower than the one expected for Pd$^{4+}$ (2.6 eV \cite{kim:pd, xray:ref}). A reduced chemical shift could be attributed to the presence of Pd$^{2+}$/Pd$^{4+}$ mixed valency. Pd$_3$O$_4$ contains a mixed valency of two Pd$^{2+}$ and one Pd$^{4+}$ ions. The peaks at 343 eV and 337.9 eV most probably are due to the presence of Pd$_3$O$_4$.

\begin{table}[h!]
\caption{The chemical shift of sample  \textsf{HT} and \textsf{ASD}, and the reference values.}
\begin{tabular}{llll}
\hline
\hline
Sample & Chem. shift:  & 3d$_{3/2}$ & 3d$_{5/2}$ \\
 & ($Pd^{2+/4+}-Pd^0$) & (eV) & (eV) \\
& & ($\pm$ 0.14 eV) & ($\pm$ 0.14 eV) \\
\hline
\textsf{HT} & Pd$^{+2}$- Pd$^0$ & 1.5 & 1.5 \\
\textsf{ASD} & Pd$^{+2}$- Pd$^0$ & 0.9 & 0.9 \\
\textsf{ASD} & Pd$^{+4}$- Pd$^0$ & 2 & 2.2 \\
\hline
reference \cite{xray:ref} & Pd$^0$-Pd$^{+2}$ & 1.2 & 1.2 \\
reference \cite{kim:pd} & Pd$^0$-Pd$^{+4}$ & 2.6 & 2.8 \\
\hline
\hline
\end{tabular}
\label{table:PdPd}
\end{table}

The Pd-3d spectra from a depth of 10-20 nm in sample \textsf{ASD} contain six compounds in addition to the Ge-LMM peaks. The two largest compounds were fitted with a pure Gaussian peak. The chemical shift between the largest compounds and the Pd$^0$ peaks is 0.9 eV for both the 3d$_{3/2}$- and 3d$_{5/2}$- peak in sample \textsf{ASD}. These peaks can probably be assigned to a Pd$^{2+}$ peak (PdO or PdO$_x$). 

Figure \ref{figure:6} shows that pure Pd is mostly found in the upper part of the nanocluster band, while the (sub)oxides were found in the lowest part of the nanocluster band, either as an oxide skin around the Pd nanocrystals and / or as pure oxide nanocrystals. This inhomogenity has most probably occurred during sputtering deposition. A small shift to a higher binding energy is observed for both Pd$^0$ (1 eV) and Pd$^{2+}$ peaks (0.5-0.7 eV) as compared to the reference values \cite{kim:pd, xray:ref}. The chemical shift between Pd$^{2+}$ and Pd$^{0}$ is lower than the reference values and lower than what was found in sample \textsf{HT}. This will be discussed in the next section.

\subsection{The nature of the Pd-3d binding energy shift}
\label{sec:shiftasd}

In a previous paper, we studied the binding energy shifts of Er$_2$O$_3$ nanoclusters in SiO$_2$, and the various factors influencing the binding energy were discussed in detail\cite{annett:erbium}. In this work we performed a similar study on Pd nanoclusters in SiO$_2$ in order to evaluate the decrease in chemical shifts found in the as deposited sample as compared to the heat treated sample. Shifts in binding energy can be expressed as

\begin{equation}
\label{eq:shift}
\Delta E_B = K \Delta q + \Delta V + \Delta \varphi - \Delta R
\end{equation}

In the above equation, $K$ is a measure of the Coulomb interaction between the valence and core electrons, and $\Delta q$ expresses changes in the valence charge. $K\Delta q$ reflects therefore charge transfer effects. $\Delta V$ is the contribution of the changes in Madelung potential. $\Delta \varphi$ contains changes in energy referencing, including variations of the sample work function and of the energy of charge compensating electrons, which may be important in the case of insulators. The first two terms in Equation \ref{eq:shift} refer to initial state effects, while the third term expresses the dependence of $\Delta E$ on energy referencing in the case of insulators. The fourth term is the contribution of the relaxation energy $R$, which is the kinetic energy gained (negative sign) when the electrons in the solid screen the photohole produced by the photoemission process; this is a final state effect.

A 1-2 eV core level shift to higher $E_B$ in the Pd-3d peak was observed in the work by Ichinohe et al. \cite{Pd:sio2}, who studied Pd clusters in SiO$_2$. The shift was attributed to final state effects as a consequence of the decrease in particle size, which is an initial state effect, and a subsequent decrease in screening. This demonstrates how the initial state influences the final state effects. In the case of nanoclusters in an insulating matrix, the core hole relaxation, screening could contain matrix contributions to some extent. Since Pd is a metal, it is characterized by a large screening efficiency. SiO$_2$, on the other hand, is an insulator and has a low screening efficiency. Therefore, assuming an external screening contribution by SiO$_2$, the screening in bulk Pd is higher than the screening in Pd nanocrystals embedded in SiO$_2$, due to the low screening contribution from the oxide. Quantum confinement effects and an increased band gap may also reduce screening since the core hole screening by the conduction band depend on the band gap. The larger the band gap is the lower the screening efficiency becomes. A reduction in core hole screening appears as an increase in binding energy.

The (sub)oxide in the as deposited samples has a decreased chemical shift compared to the heat treated samples. PdO has a higher dielectric constant than its surrounding SiO$_2$ matrix similarly to Er$_2$O$_3$ in the same matrix \cite{annett:erbium}. In accordance with the previous argumentation \cite{annett:erbium}, the screening contribution of the surrounding SiO$_2$ on the Pd and PdO clusters is expected to be small. Considering absence of energy referencing issues, initial state effects seem to have a dominant role in the increase of the binding energy of Pd nanocrystals and PdO$_x$ clusters. As for the Er$_2$O$_3$ nanoclusters \cite{annett:erbium}, charge transfer from Pd to O can lead to the creation of positive surface charges. The increased binding energy may therefore be due to initial state effects, such as charge transfer from Pd towards the interface. The smaller the nanocluster size, the higher the surface/volume ratio. Therefore interfacial transport phenomena are more enhanced. Variations in the effectiveness of charge neutralization on Pd nanocrystals with and without a PdO$_x$ skin may also account for differences in peak shifts.

\section{Conclusion}

Multilayer samples containing Pd, Ge and Si were made in order to study the nucleation, distribution, composition as well as atomic and electronic structure of Ge and Pd nanoclusters. The nanocrystals were observed by HRTEM, EDS and EFTEM imaging. Ge was observed in the form of small amorphous nanoclusters. The as deposited samples contained not only pure Pd nanocrystals, but also Pd-oxides. A 1-2 eV shift in binding energy found for the XPS Pd-3d peak of pure Pd and Pd$^{2+}$ was attributed to initial state effects arising from an increased charge transfer from Pd to O in the nanocrystals and / or to electrostatic charging. According to the combined TEM and XPS data, the Pd nanoclusters in the as deposited samples consist of Pd and PdO$_x$.

\end{document}